\documentclass[aps,amssymb,amsmath,amsfonts,twocolumn]{revtex4}
\usepackage{graphicx}
\usepackage{bm,bbm}
\usepackage{color}
\usepackage{dsfont}
\usepackage{verbatim} 

\begin{document}
\title{Universality of spectra for interacting quantum chaotic systems}
\author{
Wojciech Bruzda$^{1}$, Marek Smaczy{\'n}ski$^{{1}}$, Valerio Cappellini$^{{1}}$, Hans-J{\"u}rgen
Sommers$^{{2}}$, and Karol {\.Z}yczkowski$^{1,{3}}$\\[2ex]
{\normalsize\itshape {$^1$Mark Kac Complex Systems Research Centre, Institute of Physics,
   Jagiellonian University, ul. Reymonta 4, 30-059 Krak{\'o}w, Poland}}\\
 {\normalsize\itshape $^{{2}}$Fachbereich Physik, Universit\"{a}t Duisburg-Essen, 
   Campus Duisburg, 47048 Duisburg, Germany}\\
 {\normalsize\itshape $^{{3}}$Centrum Fizyki Teoretycznej, Polska Akademia Nauk,
   Al. Lotnik{\'o}w 32/44, 02-668 Warszawa, Poland}\\[2ex]}

 \date{{Mar. 15, 2010}}

\begin{abstract}
We analyze a model quantum dynamical system
subjected to periodic interaction with an environment,
which can describe quantum measurements.
Under the condition of strong classical chaos and 
strong decoherence due to large coupling with the measurement device,
the spectra of the evolution operator exhibit an universal behavior.
A generic spectrum consists of a single eigenvalue equal to unity, which 
corresponds to the invariant state of the system,
while all other eigenvalues  are contained in a disk in the complex plane.
Its radius
depends on the number of the Kraus measurement operators, 
and determines the speed with which an arbitrary initial state 
converges to the unique invariant state.
These spectral properties are characteristic of
an ensemble of random quantum maps, 
which in turn can be described 
by an ensemble of real random Ginibre matrices.
This will be proven in the limit of large dimension.
\end{abstract}

\maketitle
\medskip

\section{Introduction}

Time evolution of an isolated quantum system
can be described by unitary operators.
Quantum dynamics corresponds then to 
an evolution in the space of quantum pure states, 
since a given initial state $|\psi\rangle$
is mapped into another pure state
$|\psi'\rangle=U |\psi\rangle$, where $U=\exp(-{\bf i}H)$.
Here $H$ represents a Hermitian Hamiltonian of the system
and the time $t$ is set to unity.

If the underlying classical dynamics is chaotic
the Hamiltonian $H$ or the evolution operator $U$
can be mimicked by ensembles of random unitary matrices
\cite{H06,S99}. In particular,
spectral properties of an evolution operator 
of a deterministic quantum chaotic 
system coincide with predictions obtained for
the Dyson ensembles of random unitary matrices~\cite{D62}.
The symmetry properties of the system determine which
ensemble of matrices is applicable. For instance, if the physical
system in question does not possess any
time-reversal symmetry,
one uses random unitary matrices of the
circular unitary ensemble (CUE). If such a symmetry exists and
the dimension of the Hilbert space is odd
one uses symmetric unitary matrices of the circular orthogonal
ensemble (COE)~\cite{Me04}.

If the quantum system $S$ is not isolated,
but it is coupled with an environment $E$, 
its time evolution is not unitary. One needs 
then to characterize the quantum state
by a density operator $\rho$,
which is Hermitian, $\rho=\rho^{\dagger}$,
positive, $\rho \ge 0$ and normalized, Tr$\rho=1$.
Time evolution of such an open system
can be described in terms of master equations \cite{AL87},
which imply that the dynamics takes place inside the set of 
quantum mixed states.

\begin{figure}[htbp]
\centering
\includegraphics[width=0.5\textwidth]{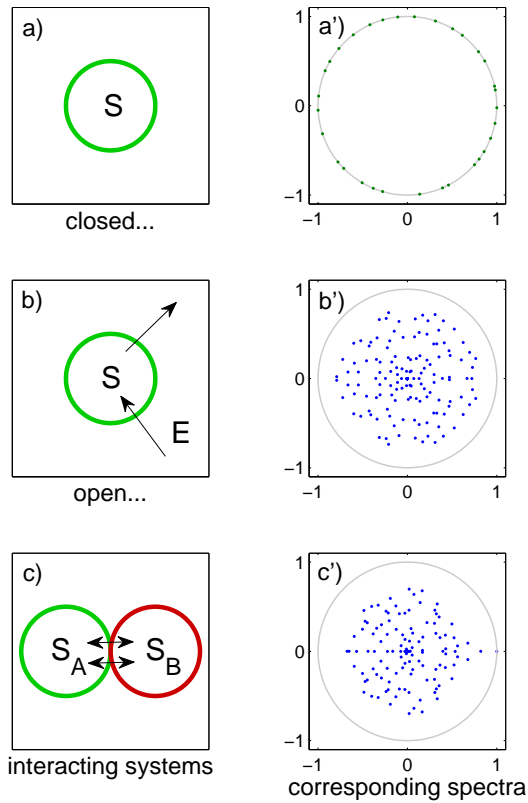}
\caption{Schematic representation of a) an isolated quantum system $S$
characterized by a Hamiltonian $H$ and a unitary evolution 
operator $U=\exp(-{\bf i}H)$;
b) open quantum system $S$. The  influence of an environment $E$ can
be described by an anti-Hermitian part of the Hamiltonian, $-{\bf i}\mu WW^{\dagger}$;
and c) interacting systems $S$ and $E$, in which the global evolution is unitary,
and the non-unitarity of the evolution of $S_A$ is due to the partial trace over the
subsystem $S_B$. Panels a'), b') and c') show exemplary spectra
of the corresponding evolution operators, which belong to the unit disk on the complex plane $z=x+iy$.}
\label{fig_schv}
\end{figure}

The coupling of the system $S$ with an environment $E$ 
can heuristically be described
by adding to the Hamiltonian an anti-Hermitian component,
$H\to H'=H-{\bf i}\mu WW^{\dagger}$,
where $W$ is an operator representing the interaction
between both systems~\cite{LW91}. The corresponding ensembles of 
non-Hermitian random matrices with spectrum supported
on the lower half of the complex plane were studied in~\cite{LSX,FS03}.
For any positive value of the coupling strength parameter 
$\mu$ the dynamics of the system is not unitary
and eigenvalues of the evolution operator
move from the unit circle inside the unit disk --
see Fig.~\ref{fig_schv}b'.
A similar situation occurs if one takes into account
dissipation in the system.
Such a dynamics of eigenvalues of a non-unitary
evolution operator in the complex plane 
was analyzed by Grobe et al.~\cite{GHS88}
and later reviewed by Haake~\cite{H06}.

Time evolution of an open quantum system can also be described
in terms of a global unitary dynamics $V$, which couples
together a system $S_A$ with an other subsystem $S_B$,
followed by averaging over the degrees of freedom
describing the auxiliary subsystem. Technically,
the image of an initial state $\rho$ of the system
 is obtained by a partial trace over the subsystem $S_B$,
$\rho'={\rm Tr}_B [V (\rho\otimes \sigma)V^{\dagger}]$,
where $\sigma$ denotes the initial state of the environment.
The map $\rho'=\Phi(\rho)$ defined in this way
is completely positive and preserves the trace,
so it is often called a {\sl quantum operation}~\cite{NC00,BZ06}.
Note that in this approach
both {\sl interacting} subsystems $S_A$ and $S_B$
are set on an equal footing. The second system,
usually referred to as an 'environment',
is in fact treated symmetrically,
and one may also consider a dual operation, 
in which the partial trace is
taken over the principal subsystem $S_A$ - compare Fig.~\ref{fig_schv}c.

A quantum map can be described by a
{\sl superoperator} $\Phi$, which
acts on the space of density operators.
If $N$ denotes the size of a density matrix $\rho$,
the superoperator is represented by a matrix 
$\Phi$ of size $N^2$. In general such a matrix is not unitary,
but it obeys a quantum analogue of
the Frobenius--Perron theorem, so its spectrum is confined to the unit disk \cite{BCSZ09}.
Spectral properties of superoperators representing 
some exemplary interacting quantum systems were analyzed
in~\cite{LPZ02,GMSS03,GMS05,PCWE09}.
It is worth to add that spectra of quantum superoperators
are already experimentally accessible: 
Weinstein et al.~\cite{WHEBSLC04} study spectra of superoperators
corresponding to an NMR realization of exemplary quantum gates.

For a quantum operation $\Phi$ there exists an invariant 
state $\omega=\Phi(\omega)$. In a generic case of a typical (random) 
operation such an invariant state is unique~\cite{BCSZ09}.
If the action of the map is repeated $n$ times
any initial quantum state $\rho$ converges to $\omega$
exponentially with the discrete time $n$.
The rate of this convergence is governed
by spectral properties of the superoperator, which can 
be characterized by the {\sl spectral gap},
defined as the difference between moduli
of the two largest eigenvalues.

The main aim of this work is to analyze spectra 
of evolution operators representing interacting quantum systems.
We demonstrate that under the condition of strong classical
chaos and strong decoherence these spectral properties are universal
and correspond to an ensemble of random operations~\cite{BCSZ09}.
In other words, we explore the link between quantum chaotic dynamics
and  ensembles of random matrices. The analysis performed earlier 
for unitary quantum dynamics~\cite{H06} (see Fig.~\ref{fig_schv}a)
is extended for a more general case of non-unitary 
time evolution of interacting quantum systems.
This problem can be described by an approach closely related to the one
used earlier to characterize quantum dissipative dynamics.
To describe spectra of such non-unitary evolution
operators Grobe et al.~\cite{GHS88} applied random matrices of
the complex Ginibre ensemble~\cite{Gi65}.

\begin{figure}[htbp]
\centering
\includegraphics[width=0.43\textwidth]{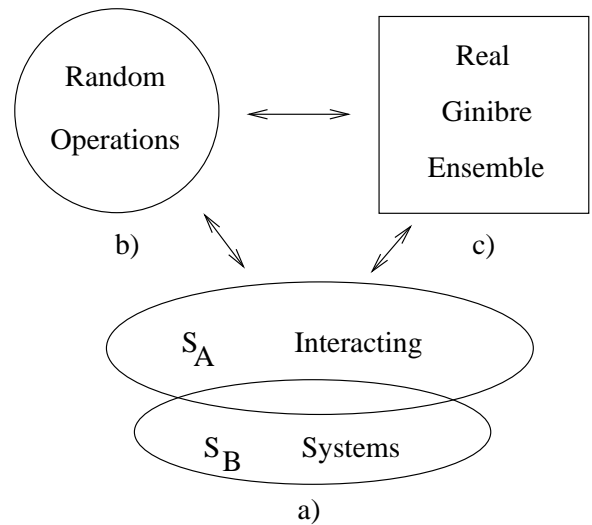}
\caption{Under the assumption of strong chaos and large decoherence
a deterministic dynamics of  a) an interacting 
 quantum system can be described by b) an ensemble
of random operations (completely positive, trace preserving maps). 
These, in turn, can be mimicked by  c) random
matrices of the real Ginibre ensemble, (which do not imply CP and TP properties).
}
\label{fig_relations}
\end{figure}

The key idea of this work
can be visualized in Fig.~\ref{fig_relations},
which shows a bridge established between 
interacting quantum systems, appropriate ensembles of random operations 
and ensembles of Ginibre matrices.
Since a superoperator describing one-step evolution
operator can be represented as a real matrix~\cite{TDV00},
we are going to apply random matrices of
{\sl the real Ginibre ensemble}~\cite{LS91,SW08}.
In particular, we investigate time evolution 
of initially random pure states in a deterministic
model of quantum baker map periodically subjected to quantum measurements,
study the speed of their convergence to the invariant state 
and compare the results with those obtained 
for an appropriate ensemble of random operations.

This paper is organized as follows. In 
section II we recall necessary definitions and
introduce several versions of a
deterministic quantum system model:
the quantum baker maps subjected to measurement process.
In Section III we introduce 
various ensembles of random maps.
In section IV the spectra and spectral gaps of 
superoperators corresponding to
baker map are analyzed and compared with the
spectra of random operations.
In section V we continue to analyze spectra of random 
quantum operations and give a proof for the real Ginibre
conjecture put forward in~\cite{BCSZ09},
and we study the fraction 
of the real eigenvalues. For large dimension $N$ this ratio is found in agreement 
with the predictions obtained for the real Ginibre ensemble.

\section{Quantum operations and spectral gap}

We shall start reviewing the necessary notions and 
definitions.
Let us define the set of quantum states ${\cal M}_N$ 
which contains all Hermitian, positive operators $\rho$ of size
$N\times N$ with trace set to unity.
A quantum linear map $\Phi$ acting on ${\cal M}_N$ is
called {\sl completely positive} (CP) if positivity of the extended map,
$(\Phi\otimes\mathbbm{1}_M)(\rho)\geqslant 0 : \forall \rho$
holds for an arbitrary dimension $M$ of the extension,
and {\sl trace preserving} (TP) if ${\rm Tr}(\Phi(\rho))={\rm Tr}(\rho).$
Any CP TP map is called quantum operation or {\sl stochastic map}.
If a quantum operation $\Phi$ preserves the identity, 
$\Phi(\mathbbm{1}/N)= \mathbbm{1}/N$, the map is called {\cal bistochastic}.

According to the dilation theorem of Stinespring~\cite{S55} any CP map
may be represented by a finite sum of $M$ Kraus operators,
\begin{equation}
\Phi(\rho)=\sum_{m=1}^M A^m\rho (A^m)^{\dagger} \ .
\label{kraus1}
\end{equation}
If the Kraus operators $A^m$ satisfy the identity resolution,
$\sum_m (A^m)^{\dagger} A^m= \mathbbm{1}_N$, the map $\Phi$ is trace preserving.
The corresponding superoperator can be expressed as a sum 
of the tensor products \cite{NC00},
\begin{equation}
\Phi=\sum_{m=1}^M A^m\otimes (A^{m})^*,
\label{Phimap}
\end{equation}
where the $^\ast$ denotes the complex conjugation.
Let $z_i$ with $i=1,\dots,N^2$ denote the spectrum of $\Phi$
ordered with respect to the moduli, 
$|z_1| \ge |z_2| \ge \dots \ge |z_{N^2}| \ge 0$.

A quantum stochastic map $\Phi$ sends the compact, 
convex set ${\cal M}_N$ of mixed density matrices into itself. Hence such a map
has a fixed point, the invariant quantum state $\omega=\Phi(\omega)$.
Thus  the spectrum of any superoperator $\Phi$
representing a quantum operation
contains an eigenvalue $z_1$ equal to unity,
while all other eigenvalues belong to the unit disk.
In the case of unitary dynamics the leading eigenvalue is degenerated,
but for a random stochastic map
 the invariant state is generically unique~\cite{BCSZ09}.
In this case any pure state, $|\psi\rangle \langle \psi|$,
converges to the equilibrium state 
$\omega$ if transformed several times by the map $\Phi$.

To analyze the rate of convergence to $\omega$ we
analyze an average trace distance to the invariant states,
\begin{equation}
 d(t)=\langle{\rm Tr}|\Phi^t(\rho_0)-\omega|\rangle_{\psi} ,
\label{avdist}
\end{equation}
where $t$ denotes the discrete time
(i.e. the number of consecutive actions of the map $\Phi$), while 
 the average is taken over the ensemble of initially random pure states, 
$\rho_0=|\psi\rangle\langle \psi|$.
In the case of a generic map an exponential convergence to equilibrium,
$d(t)=d(0)\exp(-\alpha t)$ was reported \cite{BCSZ09}.
The convergence rate depends on spectral properties of the
superoperator $\Phi$. The spectrum can be characterized by the
{\sl spectral gap}, $\gamma=1-|z_2|$,
which generically determines the convergence rate, $\alpha=-\ln(1-\gamma)$.

\section{Deterministic system: quantum baker map subjected to measurements}

The formalism of discrete quantum maps is applicable 
to describe a deterministic quantum system,
periodically interacting with an environment.
In this work we concentrate on quantum dynamical systems, 
the classical analogues of which are known to be chaotic.
Following the model of Balazs and Voros~\cite{BV89} we 
consider the unitary operator describing 
the one--step evolution model of quantum baker map, 
\begin{equation}
B=F_N^{\dagger}\left[\begin{array}{cc}F_{N/2}&0\\0&F_{N/2}\end{array}\right] .
\end{equation}
Here $F_N$ denotes the Fourier matrix of size $N$, 
$[F_{N}]_{jk}=\exp({\sf i} j k / 2\pi N)/\sqrt{N}$ 
and it is assumed that the dimension $N$ of the Hilbert space ${\cal H}_N$ is even.
 The standard quantum baker map $B$ may be generalized to represent an 
asymmetric classical map,
\begin{equation}
B_K=F_N^{\dagger}\left[\begin{array}{cc}F_{N/K}&0\\0&F_{N-N/K}\end{array}\right]
\end{equation}
where $K\ge 2$ is an integer asymmetry parameter chosen in such a way 
that the ratio $N/K$ is integer.
The standard model, obtained in the case $K=2,$
corresponds to the classically chaotic dynamics characterized by the dynamical 
entropy $H$ equal to $\ln 2$~\cite{O02}.
This system can be considered as a $2$--dimensional lift of an 
$1$--dimensional non-symmetric shift map,
\begin{equation}
f_K(x)=\left\{\begin{array}{ll}Kx/(K-1)&:x\in[0,(K-1)/K]\\&\\~Kx-K+1&: x\in((K-1)/K,1]\end{array}\right. .
\end{equation}
Chaos in  such a system can be characterized by its dynamical entropy,
equal to the mean Lyapunov exponent, averaged with respect to
the invariant measure of the classical system.
Since the uniform measure is invariant with respect to the map $f_K$,
the dynamical entropy $h$ is equal to the mean logarithm of the slope $df_K / dx$,
\begin{equation}
h(K) = \frac{1}{K}\ln K + \frac{K-1}{K}\ln\frac{K}{K-1}.
\label{entropy}
\end{equation}
The entropy is maximal in the case $K=2$, while in the limit $K \to \infty$
the entropy tends to zero. Hence the larger value of the parameter $K$ is,
the weaker chaos in the classical system becomes.

In the case of the quantum system acting on the $N$--dimensional Hilbert 
space the largest possible value of the asymmetry parameter reads $K=N$.
Thus the limiting case of the classically regular system 
cannot be obtained for any finite $N.$
The limit of vanishing dynamical entropy, $h\to 0,$ can be approached only in
the classical limit $N\to\infty$ of the quantum system.

A generalized variant of a non-unitary baker map 
introduced by Saraceno and Vallejos
described a dissipative quantum system \cite{SV96}.
In this work we study another model of non-invertible
quantum baker map analyzed  in \cite{LPZ02,ALPZ04},
which is deterministic, conserves the probability,
and is capable to describe projective measurements
or a coupling with an external subsystem.

\begin{figure}[htbp]
\centering
\includegraphics[width=0.4\textwidth]{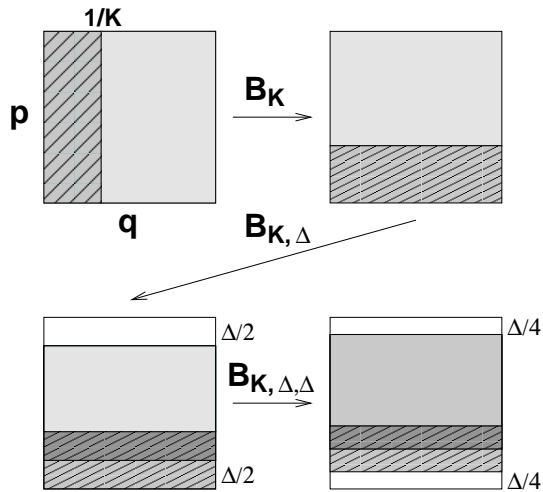}
\caption{
Sketch of the classical dynamical system acting on the torus:
b) reversible asymmetric $(K=4)$ baker map $B_K$,
c) irreversible sloppy baker map $B_{K,\Delta}$
    in which in each step the upper part of the phase space is 
    shifted down by $\Delta/2$,
d) double sloppy baker map $B_{K,\Delta,\Delta}$ in which both parts of the 
phase space are shifted vertically by $\Delta/4$.}
\label{fig_sloppy}
\end{figure}

In general there exist $M$ different outcomes of the 
measurement process and thus the map is described by a collection of
$M$ Kraus operators. The simplest  nontrivial case of $M=2$ 
corresponds to dividing of the phase space  into two parts, 
which we can choose to be the 'lower' and the 'upper' part.
Such a measurement scheme allows one to write down the quantum operation
corresponding to the 'sloppy baker map', in which both pieces of the 
classical phase space are not placed precisely one by another, but in each step
an overlap of a positive width takes place.
In the classical model the upper piece of the phase space 
is shifted down by $\Delta/2$ - see Fig. \ref{fig_sloppy}c - so the invariant 
measure lives in the rectangle of the width $(1-\Delta)$.

To represent the shift in the quantum analogue of the map
one uses a unitary translation operator $V$ 
such that $V^N=\mathbbm{1}_N$ and any momentum eigenstate $|k\rangle$ 
is shifted  by one, $V|k\rangle=|k+1\rangle$.
Hence the shift down by $\Delta/2$
is realized by the unitary operator, $V^{-N\Delta/2}$.
Thus the stochastic map describing the quantum sloppy map \cite{LPZ02} 
\begin{equation}
\Phi_{B_{K,\Delta}}(\rho) = D_b B_K\rho B_K^{\dagger} D_b^{\dagger}+ D_t B_K\rho B_K^{\dagger} D_t^{\dagger}.
\label{qbaker}
\end{equation}
consists of two Kraus operators,
 which act on the unitarily rotated state  $\rho'=B_K\rho B_K^{\dagger}$,
\begin{align}
D_b&=F_{N}^{\dagger}\left[\begin{array}{cc}\mathbbm{1}_{N/2}&0\\0&0\end{array}\right]F_N,\nonumber\\
D_t&=V^{-N\Delta/2} F_{N}^{\dagger}\left[\begin{array}{cc}0&0\\0&\mathbbm{1}_{N/2}\end{array}\right]F_N.
\label{krau2}
\end{align}
The operator $D_b$ describes the projection on the lower part of the phase space,
while the definition of the operator $D_t$ includes also
the operator representing the shift of the upper domain down by $\Delta/2$.
Observe that the parameter $\Delta$ may take any real 
value from the unit interval $[0, 1]$.
However, the case $\Delta=0$ corresponds to the baker map without the shift
but with a measurement, so it does not reduce to the standard 
unitary baker map $B_K$.

One can also consider another classical model of double sloppy map, in which both 
domains are simultaneously shifted
by $\Delta/4$ towards the center of the phase space \cite{Sm09} -- Fig. \ref{fig_sloppy}d.
To write down the corresponding quantum model $B_{K,\Delta,\Delta}$
one needs thus to modify both Kraus operators,
$D_b\to V^{N\Delta/4}D_b$ and $D_t\to V^{-N\Delta/4}D_t$.

Both variants of the model can be further generalized by allowing for a larger
number $M$ of measurement operators, represented by projectors on mutually 
orthogonal subspaces. For simplicity we assume here
that the dimensionalities of all these subspaces are equal and read $N/M$.
Varying the parameter $M$ one may thus control the degree of the 
interaction of the baker system with the environment 
and study the relation between the decoherence in the interacting quantum system
and the spectrum of the corresponding superoperator.

Increasing the asymmetry parameter $K$ one can decrease the degree of classical chaos.
To increase the degree of chaos  one may just apply the quantum baker map twice,
since the classical dynamical entropy of such a composite map is equal to $2\ln 2$. 
In general one can allow for an arbitrary number of $L$ of unitary evolutions,
and replace unitary $B$  by $B^L$. Alternatively one can say that the non-unitary measurement 
operation is performed only once every $L$ periods of the unitary evolution.
Choosing  the parameter $L$ to be of order of $N$ one can assure that
the quantum dynamics is as 'chaotic' as allowed by the quantum theory,
what can be quantitatively characterized by the quantum dynamical entropy
\cite{SZ98, BCCFV03, V07}.

Therefore the generalized model of quantum sloppy baker map we 
are going to analyze here depends on the classical asymmetry parameter $K$, 
the width of the classical shift $\Delta$,
the number of free evolutions $L$,
and the quantum parameter $M$ denoting the number of measurement operators,
\begin{equation}
\Phi_{B_{\Delta,K,L,M}}(\rho) \ =\ \sum_{m=1}^M D_m (B_K)^L \rho (B_K^{\dagger})^L D_m^{\dagger} .
\label{mapgen}
\end{equation}
Additionally, for each set of parameters of the model
one can choose the appropriate set of projection operators $D_m$
which correspond to the shift applied on one or on two parts of the classical phase space.

Note that the measurement process can also be interpreted as an interaction
with a measurement apparatus, described by an auxiliary Hilbert space 
of $M$ dimensions. Thus the model (\ref{mapgen})
represents an interacting quantum system and belongs to
 the general class of quantum operations defined by (\ref{kraus1}).
A rich structure of the model and the possibility to tune independently
several parameters of the quantum system allows us to treat this model 
as a valuable playground to investigate spectral properties of
superoperators, which represent non-unitary dynamics of interacting quantum systems.

\begin{figure}[htbp]
\centering
\includegraphics[width=0.49\textwidth]{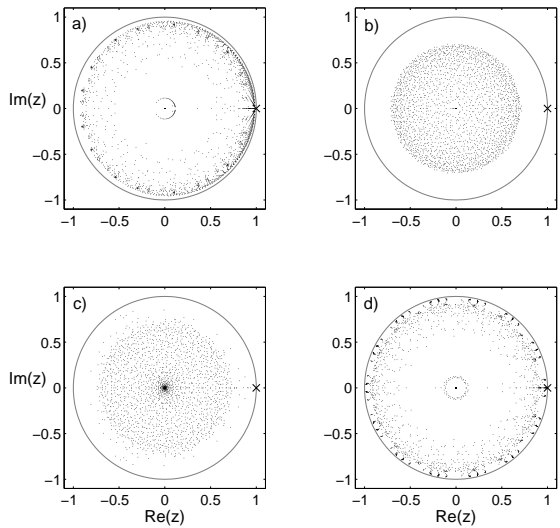}
\caption{Exemplary spectra of the evolution operator
for the  sloppy baker map for several values of the parameters of the model.
The dimension of the Hilbert space $N = 64,$ parameter $M=2$, and the shift width $\Delta=1/4$ are kept fixed.
A generic spectrum for $K=4, L=16$ is shown on panel b). The subplots a) and c)
are obtained for the cases of a weak classical chaos for $K=64, L=1$ and $K=L=32$
respectively, while the last case d) shows the spectrum for 
the double sloppy map $B_{K,\Delta,\Delta}$ for $K=64$ and $L=64$. 
}
\label{qbm64S_KLM}
\end{figure}

We constructed quantum operations representing the generalized
sloppy baker map (\ref{mapgen}) for several sets of the parameters of the model.
In each case the superoperator $\Phi$ was obtained according 
to the expression (\ref{Phimap}) and diagonalized 
to yield the complex spectrum belonging to the unit disk.

In  the case of several measurement operators, $M\ge 2$,
the quantum baker map represents a non-unitary dynamics.
Under the condition of classical chaos 
the leading eigenvalue $z_1=1$ is not degenerated
and all remaining eigenvalues are located inside
the disk of the radius equal to the modulus of the
subleading  eigenvalue $R=|z_2|$.

The spectra of the superoperator of the
generalized sloppy baker map (\ref{mapgen})
were found to depend weakly on the shift parameter $\Delta.$
However, other parameters of the model (namely $N,K,L$ and $M$)
influence properties of the spectrum considerably - 
see Fig.~\ref{qbm64S_KLM}.

As the asymmetry parameter $K$ increases the difference between the sizes
of two domains which form the classical phase space becomes larger.
In the extreme limit of $K\to N \to \infty$ the classical system
becomes only marginally chaotic,
the eigenvalues are attracted to the unit circle
and the spectral gap $\gamma=1-|z_2|$ disappears.

On the other hand, if we increase the degree of the classical chaos by increasing 
the number $L$ of unperturbed unitary evolutions, the size
of the spectral gap does not change, but the spectrum fills the complex disc
of radius $R =|z_2|$ almost uniformly.
Eventually, an increase of the number $M$ of the measurements
results in a faster decoherence in the system. This is reflected by 
an increase of the spectral gap $\gamma$. In fact the radius $R=1-\gamma$ 
of the disk supporting the spectrum decreases with $M$ as $M^{-1/2}$.
This observation  - demonstrated in  Fig.~\ref{p2nde} -
will be explained in section \ref{sec:ginib}.

\section{Ensembles of Random Operations}
\label{sec:opera}

In this section we propose three different ensembles of random stochastic maps
acting on the space ${\cal M}_N$ of mixed states of size $N$ 
with different physical interpretation. 
We assume that all unitary matrices $U$ 
used below are drawn according to the Haar measure on the unitary group
of corresponding dimension unless stated otherwise.

\medskip

{\sl 1. Environmental representation} of a random stochastic map~\cite{BCSZ09}. Choose
a random unitary matrix $U$ of composite dimension $NM$ and construct a random map as
\begin{equation}
\Phi_{\sf E}(\rho)={\rm Tr}_{\sf E}
\left[U(\rho\otimes|\nu\rangle\langle\nu|)U^{\dagger}\right].
\label{phie}
\end{equation}
It is assumed here that the environment, initially in an arbitrary pure state $|\nu\rangle$
is coupled with the system $\rho$ by a random global unitary evolution $U$.
The stochastic map is obtained by performing the partial trace
over the $M$-dimensional environment.

\medskip

{\sl 2. Random external fields} defined as a convex combination
of $M$ unitary evolutions \cite{AL87}
\begin{equation}
\Phi_{\sf R}(\rho)=\sum_{m=1}^M p_m U_m \rho U_m^{\dagger}
\label{raexfie}
\end{equation}
where $p_m$ are positive components of an arbitrary probabilistic vector of size $M$, 
$\sum_{m=1}^M p_m=1$. All unitaries $U_m\in{\sf U}(N)$ are independent 
random Haar matrices.
Random external fields form an example of bistochastic maps. They
represent the physically relevant case in which the quantum system is subjected randomly
with one of $M$ given unitary operations and can also
be interpreted as  quantum iterated function systems \cite{LSZ03}.

\medskip

{\sl 3. Projected Unitary Matrices} acting on states of a composite dimension,
$N=KM$.  All $M$ Kraus operators 
are formed by unitarily rotated projection operators, $A_m = P_m U$ for
$m=1,\dots,M$ which leads to the map
\begin{equation}
\Phi_{\sf P}(\rho)=\sum_{m=1}^M P_m U \rho U^{\dagger} P_m \ ,
\label{pum}
\end{equation}
where $U$ is a fixed random unitary matrix.
Here $P_m=P_m^{\dagger}=P_m^2$ denote projective operators
on $K$ dimensional mutually orthogonal subspaces,
 which satisfy the identity resolution, $\bigoplus_m P_m=\mathbbm{1}_N.$
 This ensemble of bistochastic maps
 corresponds to a model of deterministic quantum systems, in which
 unitary dynamics is followed by a projective measurement.

\begin{figure}[htbp]
\centering
\includegraphics[width=0.5\textwidth]{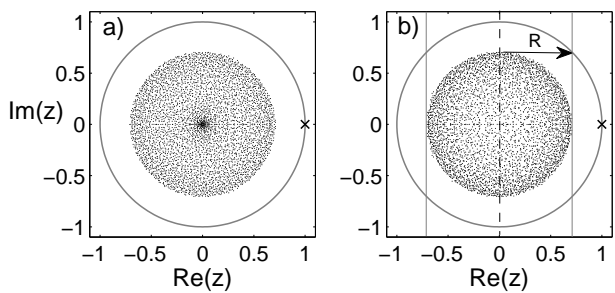}
\caption{
Spectra of superoperators corresponding to typical random maps generated
according to a) the ensemble $\Phi_{\sf E}$,
 and b) ensemble $\Phi_{\sf P}$ 
(for definitions see Sec. II). All maps act on quantum states of
size $N=64$, while the parameter of the model reads $M=2.$
Note that in all cases the spectrum is contained in the disk of radius $R=1/\sqrt{2}$
apart of the leading eigenvalue marked by '$\times$'.}
\label{rqo_exspectra}
\end{figure}

In the ensembles of random maps defined above the integer number $M\geqslant 1$
 determines the number of Kraus operators and serves as 
 the only parameter of each ensemble of random maps.
 Observe that in the special case $M=1$
the dynamics reduces to the unitary evolution, so both variants of the model are used
to describe quantum systems with or without a generalized antiunitary symmetry \cite{H06}.

As shown in \cite{BCSZ09} the flat measure in the set of stochastic matrices
is obtained for the coupling of the system with an environment of dimension $M=N^2$,
so that the Choi matrix, $D_{\Phi}:=(\Phi\otimes {\mathbbm 1})|\phi_+\rangle \langle \phi_+|$
of size $N^2$ has full rank.
Here $|\phi_+\rangle=\frac{1}{N}\sum_{i=1}^N |i,i\rangle$
represents the maximally entangled state on the 
bipartite Hilbert space, ${\cal H}_N \otimes {\cal H}_N$.
Due to the theorem of Choi the condition of complete positivity
of the map is equivalent to positivity of the Choi matrix,
\begin{equation}
\Phi {\rm \quad is \quad CP \quad}
\Leftrightarrow  \ 
D_{\Phi} \ge 0 \ .
\label{choicp}
\end{equation}

In general the discrete parameter $M$ characterizes the strength of 
the non-unitary interaction and we shall vary it from unity (unitary dynamics)
to $N^2$, which describes a generic random stochastic map.

We have generated several realizations of random maps from the ensembles
$\Phi_{\sf E}$ and $\Phi_{\sf P}$ introduced above.
Exemplary spectra of superoperators for maps pertaining 
ensembles obtained for $M=2$ are shown in Fig.~\ref{rqo_exspectra}.
In the latter case we superimposed the spectra from two realizations of the 
map $\Phi_{\sf P}$ since by construction  $N^2/M$ eigenvalues
of the superoperator are equal to zero.

In general, the spectra of random maps could be used to
describe the spectra of deterministic system  (\ref{mapgen})
under the condition of classical chaos and large decoherence.

\begin{figure}[htbp]
\centering
\includegraphics[width=0.5\textwidth]{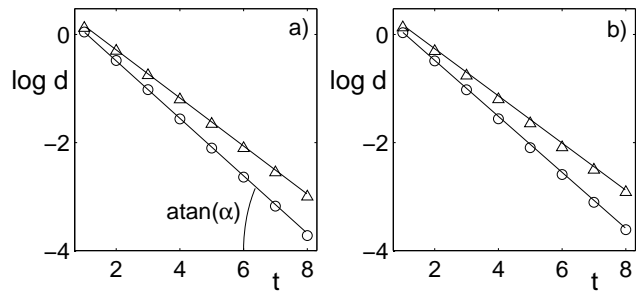}
\caption{Time dependence of the
average trace distance $d$ of random pure states to the invariant state 
for a) exemplary random quantum maps $\Phi_{\sf E}$
and b) quantum baker map with parameters: $K=4,L=16,\Delta=1/4.$
Figure drawn in a log-scale shows the exponential decay for dimension $N=24$
and parameters $M=8 (\circ)$ and $M=12$ ({\tiny $\bigtriangleup$}).
The average is taken over the set of $16$ initial projectors.
}
\label{pure_decay}
\end{figure}

Numerical results performed for various models of quantum maps 
reveal an exponential decay of the  mean trace distance  (\ref{avdist})
to the invariant state.
A comparison of such a time dependence of the mean trace distance $d$ 
for quantum baker maps and random quantum maps is shown in 
Fig.~\ref{pure_decay}.
Interestingly, for a fixed system of size $N$
the convergence rate $\alpha$ increases with the number of the measurements as 
$\alpha\sim  \frac{1}{2}\ln M$,
but it seems not to depend on the assumption,
whether the global random evolution matrix $U$
is taken from the orthogonal or unitary circular ensemble.

\vskip 0.2cm

\begin{figure}[htbp]
\centering
\includegraphics[width=0.5\textwidth]{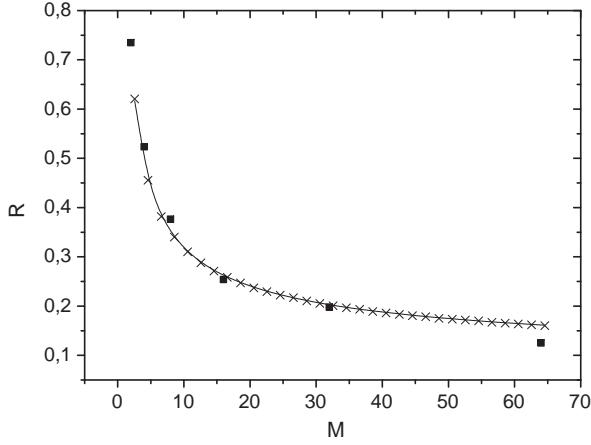}
\caption{
Modulus of the subleading eigenvalue $R=|z_2|$
as a function of the dimension $M$ of the environment for
random quantum operation $\Phi_{\sf E}$
for $N=4 \, (\times)$ and for the sloppy baker
map for $N=64, K = 4, L = 16, M = 2, \Delta=1/4 \,$ ({\tiny $\blacksquare$}).
Solid line shows the fit according to eq.~\eqref{R_sqrt_M}.}
\label{p2nde}
\end{figure}

Further numerical investigations
 confirm an expected relation between the size
 rate of the convergence of a typical quantum 
state towards the fixed point of the map and the spectral gap.
As shown in Fig.~\ref{p2nde} obtained for
random operations as well as  the generalized quantum baker map
the radius $R=1-\gamma$
of the disk in the complex plane, which contains all but the leading eigenvalue,
decreases with the number of measurements as 
\begin{equation}
R=|z_2|\sim\frac{1}{\sqrt{M}}.
\label{R_sqrt_M}
\end{equation}
In the next section we present an explanation of this 
relation based on the theory of random matrices.

\section{Quantum operations and real Ginibre ensemble}
\label{sec:ginib}

Looking at the spectrum of a random stochastic map  $\Phi_{\sf E}$
 shown in Fig. \ref{rqo_exspectra}
one can divide the entire spectrum into three parts:
i) a single eigenvalue $z_1=1$;
ii)  ${\cal N}^{\mathbbm{R}}$ real eigenvalues distributed along the real line with a density $P^{\mathbbm{R}}(x)$,
iii) remaining ${\cal N}^{\mathbbm{C}}$  complex eigenvalues $z_i$, the
distribution of which can be described by a density $P^{\mathbbm{C}}(z)$ on a complex plane.

Any density operator $\rho$ of size $N$ can be represented  
using the generalized Bloch vector representation  
\begin{equation}
\rho=\sum_{i = 0}^{N^2-1} a_i \; \lambda^i \ .
\label{eq2010}
\end{equation}
Here $\lambda^i$ denotes the generators of $\textsf{SU(}N\textsf{)}$ such
that ${\rm Tr}\left(\lambda^i\lambda^j\right)=\delta^{ij}$ and
$\lambda^0\propto 1.$ Since $\rho=\rho^{\dagger},$ $a_i\in\mathbb{R}$ for $i=0,\dots,N^2-1.$ The real
vector $[a_0,\dots,a_{N^2-1}]$ is called the generalized Bloch vector.
Thus
\begin{equation}
\Phi\left(\rho\right)=\sum_i\left(\sum_j\Phi^{ij} \; 
a_j\right)\lambda^i.
\label{eq2020}
\end{equation}

The Bloch vector can also be used to represent an arbitrary operation.
Using Kraus operators $A^m$  one 
represents the element $\Phi^{ij}$ of the superoperator $\Phi$ in a form
\begin{equation}
\Phi^{ij}
={\rm Tr}\left(\lambda^i \; \Phi\left(\lambda^j\right)\right)
={\rm Tr}\sum_m\lambda^i \; A^m \; \lambda^j \left(A^m\right)^{\dagger},\label{realrep}
\end{equation}
where $i,j=1,\dots, N^2-1$.
This square matrix of order $N^2-1$ will be called ${\sf C}$.
In a similar way we introduce the vector
$\kappa$
and find that the remaining elements of the matrix $\Phi$ do vanish,
\begin{align}
\Phi^{i0}&={\rm Tr}\sum_m\lambda^i A^m\lambda^0\left(A^m\right)^{\dagger}
          ={\rm Tr} \; \lambda^i\lambda^0\sum_m
          a_m\left(A^m\right)^{\dagger}\equiv (\kappa)_i
\label{eq2040}
\\
\Phi^{0j}&={\rm Tr}\sum_m\lambda^0 A^m\lambda^j\left(A^m\right)^{\dagger}
          ={\rm Tr} \; \lambda^0\lambda^j = \delta^{0j} \ .
\label{eq2030}\\
\end{align}
Hence the superoperator $\Phi$ can be represented as a real asymmetric matrix
\begin{equation}
\Phi_{ij}=\left[
\begin{array}{ll}
  1 & 0 \\
  \kappa&{\sf C}
\end{array}
\right],
\label{super2}
\end{equation}
where the $N':=N^2-1$ dimensional vector $\kappa$ represents a translation 
vector while
the $N'\times N'$ real matrix ${\sf C}$ is a real contraction~\cite{BCSZ09}.
Thus the spectrum of $\Phi$ consists of the leading eigenvalue, equal to unity,
and the spectrum of ${\sf C}$.
Note that the complex eigenvalues of the real matrix ${\sf C}$
appear in conjugated pairs, $z$ and $\bar z$, which is a consequence
of the fact, that the map $\Phi$ sends the set of Hermitian operators into itself,
so as seen above the superoperator can be represented by a real matrix \cite{TDV00}.
Since a map acting on states of size $N$ is represented by a 
superoperator of dimension $N^2$ the following normalization relation holds,
  $1+{\cal N}^{\mathbbm{R}}+{\cal N}^{\mathbbm{C}} = N^2$.

In the case that ${\sf C}$ has only real eigenvalues one
can bring ${\sf C}$ by an orthogonal transformation ${\cal O}$
to lower triangular form
\begin{equation}
{\sf C}={\cal O}\left(\Xi+\Lambda\right){\cal O}^{-1}
\end{equation}
where $\Lambda={\sf diag}(z_1,\dots,z_{N'})$ while $\Xi$ has elements
only below the diagonal. Thus
\begin{align}
d{\sf C}={\cal O}[{\cal O}^{-1}d{\cal O}\left(\Xi+\Lambda\right)&-
\left(\Xi+\Lambda\right){\cal O}^{-1}d{\cal O}+\nonumber\\
&+d\Xi+d\Lambda]{\cal O}^{-1}.
\end{align}
Hence the measure $D{\sf C}$ is given by
\begin{equation}
D{\sf C}=\left|\prod_{i<j}(z_i-z_j)\prod_k dz_k\prod_{i<j}({\cal O}^{-1}d{\cal O})_{ij}
\prod_{j<i}d\Xi_{ij}\right|
\end{equation}
where the Vandermonde determinant is the Jacobian of the transformation
from $({\cal O}^{-1}d{\cal O})_{ij}\Lambda_j$ to $({\cal O}^{-1}d{\cal O})_{ij}.$
Thus the measure $d\mu(\Lambda)$ has the form
\begin{equation}
D\mu(\lambda)=\left|\prod_{i<j}(z_i-z_j)\prod_k dz_k\right|\overline{\Theta(D_{\Phi}\geqslant 0)}
\end{equation}
where in the last factor the positivity conditions of the corresponding Choi
matrix is averaged by integrated over the measure $D\kappa D\Xi\prod{\cal O}^{-1}d{\cal O}.$
This factor is expected to be a smooth function of the eigenvalues $z_1,\dots,z_{N'}.$
In the case that ${\sf C}$ has a certain number of complex conjugate eigenvalues
$D\mu(\Lambda)$ is of similar form, but the product of
differentials $dz_k$ has to be interpreted as exterior product~\cite{SW08}.
It turns out that for large dimension $N'$ the measure $d\mu(\Lambda)$ is
given by the real Ginibre ensemble with
the bulk of eigenvalues inside a certain disk in the complex
plane. To prove this let us go back to the matrix
representation of $\Phi$ in terms of $M$ Kraus operators $A^m,m=1,\dots,M.$
Then
\begin{equation}
\Phi(\rho)_{ij}=\sum_{m=1}^M A^m_{ik}\rho_{kl}(A^{m}_{jl})^{\ast}=\sum_{kl}\Phi_{ij,kl} \; \rho_{kl}
\end{equation}
where $A^m_{ik}$ are the matrix elements of Kraus operators and $^\ast$ denotes
the complex conjugation; $i,j,k,l=1,\dots,N.$ The Kraus operators obey
\begin{equation}
\sum_{m=1}^M\sum_{i=1}^N A^m_{ik} (A^{m}_{il})^{\ast}=\delta_{kl}
\end{equation} 
thus it is natural to assume that $A^m_{ik}$ represent $N$ columns
of a matrix $U$ drawn from a circular unitary ensemble of dimension
$NM$ i.e. $U\in{\sf U}(NM).$ Using formulas by Mello~\cite{Mello} for
the first four moments of ${\sf U}(NM)$ we are able
to find 
exactly the first two moments of matrix elements
$\Phi_{ij,kl}$.
For example
for $U\in{\sf U}(N):\langle U_{b\beta} U^{\ast}_{a\alpha}\rangle=\delta_{ab}\delta_{\alpha\beta}/N.$
Here $\langle\cdots\rangle$ means the averge over the unitary group.
This implies here:
\begin{equation}
\langle\Phi_{ij,kl}\rangle=\sum_{m=1}^M\langle A^m_{ik} (A^{m}_{jl})^{\ast}\rangle
=\frac{1}{N}\delta_{ji}\delta_{lk}.
\end{equation}
In this way we can derive the exact second moments:
\begin{align}
\langle&\Phi_{ij,kl}\Phi^{\ast}_{\bar{i}\bar{j},\bar{k}\bar{l}}\rangle=\nonumber\\
&\frac{1}{(NM)^2-1}\left(M^2\delta_{ij}\delta_{\bar{i}\bar{j}}\delta_{kl}\delta_{\bar{k}\bar{l}}
+M\delta_{i\bar{i}}\delta_{j\bar{j}}\delta_{k\bar{k}}\delta_{l\bar{l}}\right)\nonumber\\
-&\frac{1}{NM}\frac{1}{(NM)^2-1}\left(M^2\delta_{ij}\delta_{\bar{i}\bar{j}}\delta_{k\bar{k}}\delta_{l\bar{l}}
+M\delta_{i\bar{i}}\delta_{j\bar{j}}\delta_{kl}\delta_{\bar{k}\bar{l}}\right).
\end{align}
We see that in the limit of large $N$ the first two cumulants are
identical to those of the Gaussian distribution (with variance $1/(N^2M)$)
\begin{equation}
P(\Phi)\propto
 \exp\left(NM\sum_{ik}\Phi_{ii,kk}-\frac{N^2M}{2}\sum_{ijkl}|\Phi_{ij,kl}|^2\right).
\end{equation}
The factor $1/2$ is due to the symmetry property
$\Phi_{ij,kl}=\Phi^*_{ji,lk}$. We will argue below that for large $M$ in
addition the higher cumulants can be neglected.
Hence the superoperator $\Phi$  associated with a random map
can be described (up to the one eigenvalue $1$) by the 
real Ginibre ensemble with eigenvalues inside a disk of radius $1/\sqrt{M},$ where $M$ is the
number of random Kraus operators defining $\Phi.$ This can also
be seen by going back to the real matrix representation~\eqref{realrep}-\eqref{super2}.

Let us now argue that for large $M$ we can neglect higher cumulants. First
of all for large $N$ the elements $A^m_{ik},$ forming a minor of $U \in {\sf U}(NM),$ are
essentially independent Gaussian variables with zero mean and variance $1/NM$. Thus as
a consequence of the central limit theorem for large $M$ $\Phi_{ij,kl}$ as sum of $M$
essentially independent identically distributed variables is again
Gaussian with variance $M/(NM)^2=1/N^2M$. Also in the bulk the different
matrix elements of $\Phi$ are independent. The average of $\Phi_{ij,kl}$ is
given by $\delta_{ij}\delta_{kl}/N$.

\begin{figure}[htbp]
\centering
\includegraphics[width=0.46\textwidth]{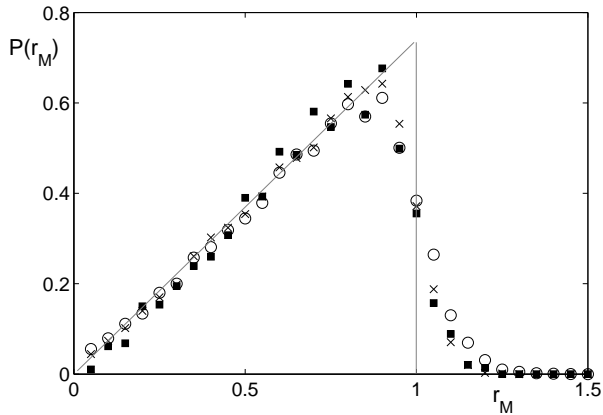}
\caption{Radial density of complex eigenvalues of spectra
 of superoperators corresponding to a deterministic model of
 sloppy baker map $(\times)$, projective random operations $(\circ)$
 and the spectra of real random matrices of the Ginibre ensemble ({\tiny $\blacksquare$}).
 The size of each matrix is $N^2=32^2$, the number of
 Kraus operators is $M=16$ so the density is shown as a function 
 of the rescaled radius $r_M:=r\sqrt{M}=4r$.
The tail of the distribution beyond the point $r_M=1$ (equivalent to $r=1/4$)
does not violate therefore the quantum analogue of the Frobenius-Perron theorem.}

\label{ep_joint}
\end{figure}

To investigate the density of complex eigenvalues $z=x+iy$ 
of the superoperator $\Phi$ we analyzed their
radial probability distribution $P(r)$, where $r=|z|$.
The real eigenvalues are taken into account for this statistics.
Fig. \ref{ep_joint} shows a comparison of numerical data
obtained for several realizations of quantum baker map,
projective random operations and real random matrices
pertaining to the Ginibre ensemble.
The data are represented in the rescaled variable
$r_M=r\sqrt{M}$ so that the radius of the disk 
of eigenvalues is set to unity.
In all three cases displayed in the figure the radial density 
grows linearly, which corresponds to the flat distribution
of eigenvalues inside the complex disk,
in agreement with the predictions of the Ginibre ensemble.
These results obtained for $N=32$ show a smooth
transition of the density in the vicinity of the boundary of the disk at $r_M=1$,
which becomes more abrupt for larger $N$.
In the asymptotic case $N\to \infty$,
the density of rescaled  eigenvalues is described by the {\sl circular law}
of Girko, 
\begin{equation}
P^{\mathbbm{C}}(z) \sim \Theta(1-|z|).
\end{equation}
derived for complex Ginibre matrices.

The spectra of real random Ginibre matrices
display a more subtle structure.
A finite fraction of all eigenvalues are real,
in analogy to the 
mean number of real roots of a real polynomial \cite{BBL92,AK07}.
Real eigenvalues of a real Ginibre matrix
cover the real axis with a constant density.
Furthermore, for large dimension the density of complex eigenvalues 
is known to be asymptotically
constant in the complex disk except for a small region near the real axis~\cite{LS91}.

\begin{figure}[htbp]
\centering
\includegraphics[width=0.5\textwidth]{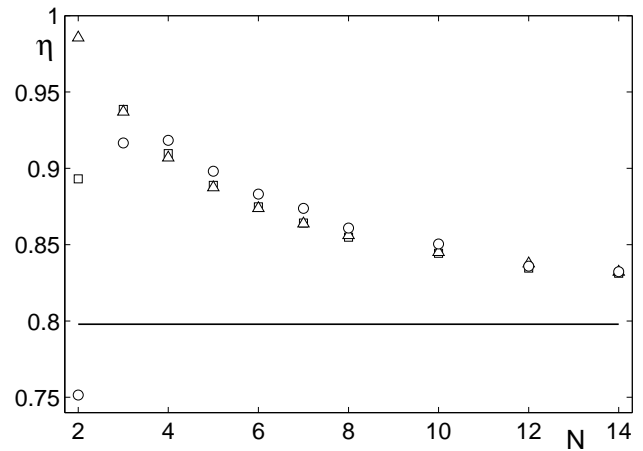}
\caption{Rescaled ratio of the real eigenvalues 
$\eta$ 
of the superoperator
for random operations with $M=N^2$
({\tiny $\square$}), $M=N$ $(\circ)$
and real Ginibre matrices ({\tiny $\bigtriangleup$}) 
as a function of the matrix size $N$.
Solid horizontal line  at $\sqrt{2/ \pi}$ 
represents the asymptotic value of the normalized ratio
implied by (\ref{ed_eq}).}
\label{real_per}
\end{figure}

To check for what random operations
these effects can be observed in the spectrum of the superoperator
we analyzed the average number 
$\langle {\cal N}^{\mathbbm{R}}\rangle_{\Phi}$
of real eigenvalues of the superoperator $\Phi$.
For any realization of $\Phi$ we have
${\cal N}^{\mathbbm{R}}=\#{\sf REAL} / (N^2-1)$
where $\#{\sf REAL}$ is the number of real eigenvalues 
of the real matrix of size $N'=N^2-1$.
These data are compared with predictions for the 
real Ginibre ensemble,
hereafter denoted by $\langle {\cal N}^{\mathbbm{R}}\rangle_{\sf RG}$.
The following expression
for the mean number of real eigenvalues of a real Ginibre matrix of size $N^2-1$
was derived in \cite{EK95,E97,So07}

\begin{align}
\langle {\cal N}^{\mathbbm{R}}\rangle_{\sf RG}&=
1+\frac{\sqrt{2}}{\pi}\int_0^1
\frac{t^{1/2}(1-t^{N^2-2})dt}{(1-t)^{3/2}(1+t)}\nonumber\\
&\simeq\sqrt{\frac{2}{\pi}}\sqrt{N^2-1}
{\rm \quad as \quad}
 N\to\infty \ .
\label{ed_eq}
\end{align}
These analytical results suggest to introduce a rescaled
ratio
\begin{equation}
\eta :=\frac{\langle {\cal N}^{\mathbbm{R}}\rangle_{\Phi}}{\sqrt{N^2-1}}
\label{etadef}
\end{equation}
to make easier a comparison of data obtained for various
systems of size $N$.
Numerical results presented in  Fig. \ref{real_per}
show that the superoperators associated with random maps
are characterized by a non-zero fraction of real eigenvalues.
In the case of strong interaction with the ancilla of the size $M=N^2$
the dynamical matrix $D_{\Phi}$ has full rank and the rescaled
fraction of real eigenvalues of $\Phi$ coincides with the
prediction for the real Ginibre ensemble.

\begin{figure}[htbp]
\centering
\includegraphics[width=0.5\textwidth]{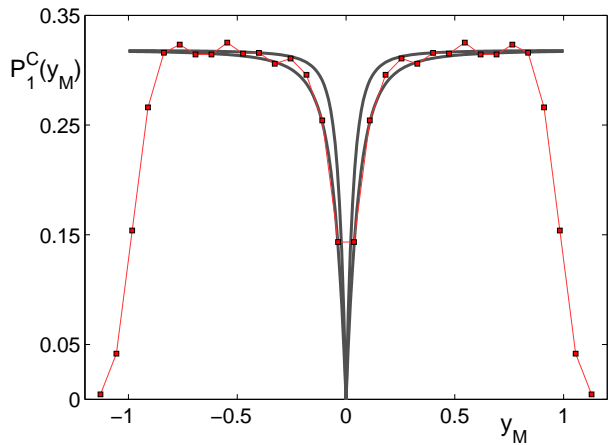}
\caption{
Numerical data for the density $P_1^{\mathbbm{C}}(y_M)$
of complex eigenvalues along the rescaled imaginary axis
obtained for an ensemble of random operations $\Phi_{\sf E}$ of dimension
$N=8$ and with parameter $M=N^2=64$ ({\tiny $\blacksquare$})
are compared with lower and upper bounds (\ref{bounds}) 
obtained for small $|y|$ from the real Ginibre ensemble
and represented by thick lines.}
\label{ginoe_cusps}
\end{figure}

To demonstrate further spectral features characteristic of the real Ginibre ensemble
we analyzed spectra of superoperators and 
investigated the cross-section of the probability distribution $P(z)$
along the imaginary axis. Numerical data of this distribution
denoted as  $P_1^{\mathbbm{C}}(y)$ obtained for an ensemble of random maps
 $\Phi_{\sf E}$ acting on the states of size $N=3$
 are shown in Fig. \ref{ginoe_cusps}.

In order to compare these data with predictions of the real Ginibre ensemble
we need to assure a suitable normalization.
Let us rescale the imaginary axis as $y\mapsto y_M=\sqrt{M} y$, 
so that the rescaled formula (25) of \cite{So07} takes the form
\begin{align}
P_1^{\mathbbm{C}}(y_M)&:=R_1^{\mathbbm{C}}(\sqrt{M} y)\nonumber\\
&\simeq \sqrt{\frac{2M}{\pi}}\exp\left(2My^2\right)|y|{\sf erfc}(|y|\sqrt{2M}).
\label{eq_erfc}
\end{align}
Making use of the standard estimations 
\begin{equation}
\frac{1}{x+\sqrt{x^2+2}}<\exp(x^2)\int_x^{\infty}\exp(-t^2)dt\leqslant\frac{1}{x+\sqrt{x^2+\frac{4}{\pi}}}
\end{equation}
(see formula (7.1.13) at page 298 of Abramowitz and Stegun \cite{AS72}),
and the definition of the complementary error function ${\sf erfc}(z)$, 
one obtains from  eq.~(\ref{eq_erfc}) an explicit form 
for lower and upper bounds for the rescaled distribution
in the vicinity of the real axis 
\begin{equation}
\frac{1}{\pi}\frac{2}{1+\sqrt{1+\frac{1}{My^2}}}
\leqslant  P_1^{\mathbbm{C}}(y_M) \leqslant
\frac{1}{\pi}\frac{2}{1+\sqrt{1+\frac{2}{\pi} \frac{1}{M y^2}}}.
\label{bounds}
\end{equation}
As shown in Fig.~\ref{ginoe_cusps} these bounds are rather precise and 
describe well the numerically observed density 
$P_1^{\mathbbm{C}}(y)$ of complex eigenvalues 
of the superoperators along the imaginary axis.

\section{Concluding remarks}

In this work we analyzed spectra of non-unitary evolution operators
describing exemplary quantum chaotic systems
and the time evolution of initially pure states.
We have chosen to work with a generalized model
of quantum baker map subjected to measurements \cite{LPZ02,Sm09},
which allows one to control the degree of classical chaos
and the strength of the interaction with the environment.
The size of the quantum effects,
proportional to the ratio of the Planck constant to the typical action 
in the system, is controlled by the size $N$ of the Hilbert space
used to describe the quantum system. The classical limit of the
quantum model corresponds to the limit $N\to \infty$.

Due to a quantum analogue of the Frobenius--Perron theorem the 
evolution operator has at least one eigenvalue equal to unity, 
while all other eigenvalues are contained in the unit disk in the complex plane.
In a generic case the leading eigenvalue is not degenerated and
the corresponding eigenstate represents the unique quantum state
invariant with respect to the evolution operator.

Investigating the time evolution of initially random pure states
we found out that in a generic case they converge
exponentially fast to the invariant state.
 The rate of this relaxation to the equilibrium 
depends on the size of the spectral gap, 
equal to the difference between the moduli of the
first and the second eigenvalues of the evolution operator.
In particular, the relaxation rate $\alpha$ was found to depend on the
number of measurement operators $M$ as $\frac{1}{2}\ln M$.

Spectral properties of evolution operators of deterministic 
quantum systems interacting with the environment were compared
with spectra of suitably defined ensembles of random matrices.
Note that an idea to apply random matrices to model evolution
operators of open deterministic quantum systems was put forward by 
Pep{\l}owski and Haake \cite{PH93}, but random maps
used therein are not necessarily completely positive.
This property is by construction fulfilled by the ensemble
of random stochastic maps introduced in ~\cite{BCSZ09}
and by two other ensembles of random bistochastic maps used
in this work.

In general, the spectra of non-unitary operators
corresponding to quantum deterministic systems
display a wide variety of structures, which depend
on classical parameters as the degree of chaos
of the corresponding classical system 
characterized quantitatively e.g. by its dynamical entropy.
The spectra depend also on quantum parameters 
as the dimension of the Hilbert space
and the character of the interaction with the environment, which 
governs the strength of the decoherence effects.
However, under an assumption of strong classical chaos
and a uniform coupling of the system analyzed with all the states of the 
$M$--dimensional environment the spectral properties the corresponding  evolution
 become {\sl universal}:
The spectrum consists of a leading eigenvalue equal to unity,
while all other eigenvalues cover the complex disk of radius $R=1/\sqrt{M}$.

In the asymptotic limit $N\to \infty$ the density of complex eigenvalues becomes
uniform in the disk, besides the region close to the real axis.
As the size of the environment $M$ is equal to $N^2$,
which implies strong decoherence, the dynamical matrix $D_{\Phi}$ describing the
quantum map $\Phi$ has full rank, so it can be considered as generic.
In this very case the spectral statistics of this region of the complex spectrum
of the superoperator and the fraction of its real eigenvalues
coincides with predictions of the {\sl real Ginibre ensemble}, a proof
of which is given in this work.

Acknowledgements.
It is a pleasure to thank R. Alicki
for several fruitful discussions on various versions
of the model of quantum interacting baker map.

Financial support by the Transregio-12 project 
der Deutschen Forschungsgemeinschaft and
the special grant number DFG-SFB/38/2007 of
Polish Ministry of Science and Higher Education
is gratefully acknowledged.

\end{document}